\documentclass[preprint,showpacs,showkeys,preprintnumbers,amsmath,amssymb,superscriptaddress]{revtex4}

\usepackage{graphicx}
\usepackage{dcolumn}
\usepackage{bm}
\usepackage[latin1]{inputenc}

\def\Ubc{\hat{\mathbf{U}}_{bc}}
\def\Uab{\hat{\mathbf{U}}_{ab}}

\def\r{\vec{r}}
\def\R{\mathbf{R}}
\def\e{\hat{e}}
\def\n{\hat{n}}
\def\J{\vec{\mathbf{J}}}

\begin{document}

\author{R. Rossi Jr.}
\affiliation{Departamento de F\'{\i}sica, Instituto de Ciências
Exatas, Universidade Federal de Minas Gerais, C.P. 702, 30161-970,
Belo Horizonte, MG, Brazil}

\author{A. R. Bosco de Magalh\~{a}es}
\affiliation{Departamento de F\'{\i}sica e Matem\'{a}tica,\\  Centro
Federal de Educa\c{c}\~{a}o Tecnol\'{o}gica de Minas Gerais, Av.
Amazonas 7675, CEP 30510-000, Belo Horizonte, MG, Brazil}

\author{J.G. Peixoto de Faria}
\affiliation{Departamento de F\'{\i}sica e Matem\'{a}tica,\\  Centro
Federal de Educa\c{c}\~{a}o Tecnol\'{o}gica de Minas Gerais, Av.
Amazonas 7675, CEP 30510-000, Belo Horizonte, MG, Brazil}

\author{M. C. Nemes }
\affiliation{Departamento de F\'{\i}sica, Instituto de Ciências
Exatas, Universidade Federal de Minas Gerais, C.P. 702, 30161-970,
Belo Horizonte, MG, Brazil}

\title{Control of state and state entanglement with a single auxiliary subsystem}

\begin{abstract}
We present a strategy to control the evolution of a quantum system.
The novel aspect of this protocol is the use of a \emph{single
auxiliary subsystem}. Two applications are given, one which allows
for state preservation and another which controls the degree of
entanglement of a given initial state.

\end{abstract}

 \pacs{03.65.Xp, 03.65.Ud, 03.67.Bg}

\maketitle

\section{Introduction}

Protocols for control and manipulation of quantum systems are
essential for the development of quantum information theory
\cite{art1}. Advances of this area can provide proper tools to avoid
decoherence and to conduct quantum evolution to desirable results.
Some examples of those strategies are: Quantum Zeno Effect (QZE)
\cite{art2,art3,art4}, Super Zeno Effect \cite{art5}, strong
continuous coupling \cite{art6,art7}, Bang-Bang control
\cite{art8,art9}, etc.

In the present contribution we show a new possibility to control the
evolution of a quantum system through successive interactions with a
single auxiliary subsystem. These interactions are described as
unitary evolutions (in a finite time period), that can stop or
inhibit the evolution of the system of interest.

To make our results concrete, we present the protocol based on two
examples. In the first one, we study a system of interest composed
by two coupled qubits ($S_{b}$ and $S_{c}$) sharing one excitation.
A third two level system ($S_{a}$) is the auxiliary subsystem
responsible for the control. Interactions between $S_{b}$ and
$S_{a}$ are inserted in the evolution of the system $S_{b}-S_{c}$
and control its dynamics. The quantity of such interactions, as well
as their duration are parameters that allows for several forms of
control. We performed an analytical calculation for the state vector
of the global system after $N$ interactions with $S_{a}$. The
Hamiltonian form of this system permits the mapping of the global
evolution ($S_{a}-S_{b}-S_{c}$) in the real euclidian subspace,
suggesting a geometrical interpretation for this dynamics. We also
pointed out the differences between the present dynamics and the
QZE.

In the second example we show how to control the entanglement
dynamics presented in Ref. \cite{art10}, through interactions with a
single auxiliary subsystem. In Ref. \cite{art10}, two initially
entangled atoms undergo different time evolutions. One of them
interacts with an electromagnetic mode in a cavity and the other one
evolves freely. The dynamics is nondissipative and the entanglement
oscillates. The introduction of an auxiliary subsystem, that
interacts with the atom in the cavity, allows for the control of
entanglement dynamics. An empirical implementation for this process
may be realized with the experimental setup used in Ref.
\cite{art11}, where a two level atom interacts with two
electromagnetic modes preserved in the same microwave cavity. In
such empirical implementation, the two level atom and one of the
modes in the cavity ($M_{a}$) compose the system of interest, and
the second mode ($M_{b}$) acts as an auxiliary subsystem.

\section{Two qubits dynamics}

Let us consider the system of interest composed by two coupled
qubits ($S_{b}-S_{c}$) and another qubit ($S_{a}$) as an auxiliary
system. The hamiltonian that governs the interaction between $S_{b}$
and $S_{c}$ is given by:

\begin{equation}
H_{bc}= \epsilon_{a}|1_{a}\rangle\langle 1_{a}| +
\epsilon_{b}|1_{b}\rangle\langle 1_{b}| +
\epsilon_{c}|1_{c}\rangle\langle 1_{c}| +I_{a}\otimes\hbar
G_{bc}(\sigma^{b}_{-}\sigma^{c}_{+} + \sigma^{b}_{+}\sigma^{c}_{-}),
\end{equation}
where $\sigma_{+}=|1\rangle\langle 0|$, $\sigma_{-}=|0\rangle\langle
1|$, $G_{bc}$ is the coupling coefficient, $I_{a}$ is the identity
matrix on the subsystem $S_{a}$. The coefficients $\epsilon_{a}$,
$\epsilon_{b}$ and $\epsilon_{c}$ are the eigenvalues of the free
hamiltonian.

The goal is to control the dynamics in subsystem $S_{b}-S_{c}$
through interactions between the auxiliary qubit $S_{a}$ and
$S_{b}$. The hamiltonian for these auxiliary interactions is

\begin{equation}
H_{ab}= \epsilon_{a}|1_{a}\rangle\langle 1_{a}| +
\epsilon_{b}|1_{b}\rangle\langle 1_{b}| +
\epsilon_{c}|1_{c}\rangle\langle 1_{c}| +\hbar
G_{ab}(\sigma^{a}_{-}\sigma^{b}_{+} +
\sigma^{a}_{+}\sigma^{b}_{-})\otimes I_{c},
\end{equation}
where $I_{c}$ is the identity matrix on $S_{c}$ and $G_{ab}$ is the
coupling coefficient between $S_{a}$ and $S_{b}$.

Suppose that $S_{a}-S_{b}-S_{c}$ share one excitation and
$\epsilon_{a}=\epsilon_{b}=\epsilon_{c}$. Since the operators
$H_{bc}$ and $H_{ab}$ preserve the excitation number, we can write
in one excitation subspace the time evolution operators as

\begin{equation}
  \Ubc(\theta) = \left[
    \begin{array}{ccc}
      1 & 0 & 0 \\
      0 & \cos\theta & -i\sin\theta \\
      0 & -i\sin\theta & \cos\theta
    \end{array} \right],\label{Utheta}
\end{equation}
and
\begin{equation}
  \Uab(\phi) = \left[
    \begin{array}{ccc}
      \cos\phi & -i\sin\phi & 0 \\
      -i\sin\phi & \cos\phi  & 0 \\
          0    &    0      & 1
    \end{array} \right],\label{Uphi}
\end{equation}
where $\theta=G_{bc}t_{bc}$, $\phi=G_{ab}t_{ab}$, $t_{bc}$
($t_{ab}$) is the interaction time between $S_{b}$ and $S_{c}$
($S_{a}$ and $S_{b}$).

The control of $S_{b}-S_{c}$ dynamics is induced by unitary
operators ($\Uab$) inserted $N$ times in the free evolution of
$S_{b}-S_{c}$. The number of interventions and the durations of each
one are the parameters that specify the control. The general
expression for the state vector of the global system submitted to
this control is given by:

\begin{equation}
  \left| \psi_N\right\rangle = \left(\Uab(\phi)\Ubc(\theta)\right)^N
  \left|\psi(0)
  \right\rangle,
  \label{csiN}
\end{equation}
where the time evolution of $S_{b}-S_{c}$ was divided by $N$
interactions with the auxiliary subsystem.

In the appendix we calculate the vector state $\left|
\psi_N\right\rangle$ for a general initial state. Now, let us
consider the initial state $\left|\psi(0)
  \right\rangle= \left|0_{a}\right\rangle\left|0_{b},1_{c}
  \right\rangle$ which goes through a quantum transition when submitted to the time evolution
  $\Ubc(\frac{\pi}{2})\left|\psi(0)  \right\rangle=\left|0_{a}\right\rangle\left|1_{b},0_{c}
  \right\rangle$. We show the inhibition of this transition through a sequence of unitary interactions with
  the single auxiliary subsystem.

 As it is shown in the appendix, the global evolution of this system can be mapped on
 $R^{3}$, therefore we may represent a sequence of $N$ interactions with the auxiliary subsystem as:

\begin{equation}
    \r_{N}=\left[\R_3\left(\phi\right)\R_1\left(-\frac{\pi}{2N}\right)\right]^N \r(0)=\left[
    \begin{array}{c}
     ac\left(1-\cos N\varphi\right)+b\sin N\varphi\\
     bc\left(1-\cos N\varphi\right)-a\sin N\varphi\\
     \left(1-c^2\right)\cos N\varphi + c^2
    \end{array}\right],
\end{equation}
where $\phi\neq 2\pi$ and $\r(0)=\left[
    \begin{array}{c}
      0 \\
      0\\
      1
    \end{array} \right]$.
Taking the limit $N\rightarrow \infty$ we have

\begin{equation}
\lim_{N\rightarrow \infty}\r_{N}=\r(0),
\end{equation}
as $a\rightarrow 0$, $b\rightarrow 0$ and $c\rightarrow 1$.

To give a geometrical interpretation of this effect consider a
vector $\r$ in the euclidian subspace. In the present dynamics, the
rotations $\R_1\left(-\theta\right)$ around the axis $Ox$ are
clockwise. Therefore, when $\r$ has a positive $y$ component, the
rotations $\R_1\left(-\theta\right)$ will reduce the $z$ projection
of the vector, but when $\r$ has a negative $y$ component the
rotation $\R_1\left(-\theta\right)$ will do the opposite, tending to
compensate the previous decrease. The rotations
$\R_3\left(\phi\right)$ move the vector through the subspaces where
the $y$ component is positive and where it is negative. Therefore,
when we study the dynamics of $\r$ inserted by rotations
$\R_3\left(\phi\right)$ we notice that the decreasing of the $z$
projection, induced by $\R_1\left(-\theta\right)$ when $y>0$, is
compensated by the increasing of the same projection, also induced
by $\R_1\left(-\theta\right)$, but when $y<0$. Choosing the angle of
the rotations $\R_3\left(\phi\right)$ and the number of
interventions $N$ it is possible to preserve the projection $z$ of
$\r$ or even freeze the dynamics of $\r$. In Fig.1 we explicitate
another geometric point of view for the effect: the net effect of
$\R_{1}\left( -\theta \right)$ and $\R_{3}\left( \phi \right)$
concerns a rotation around the vector $\hat{n}$, when the relation
$\theta /\phi $ decreases, $\hat{n}$ gets closer to $\e_3$ and the
state will be always closer to the initial one.

\begin{figure}[h]
\centering
%\hspace{-3.5cm}
  \includegraphics[scale=0.4]{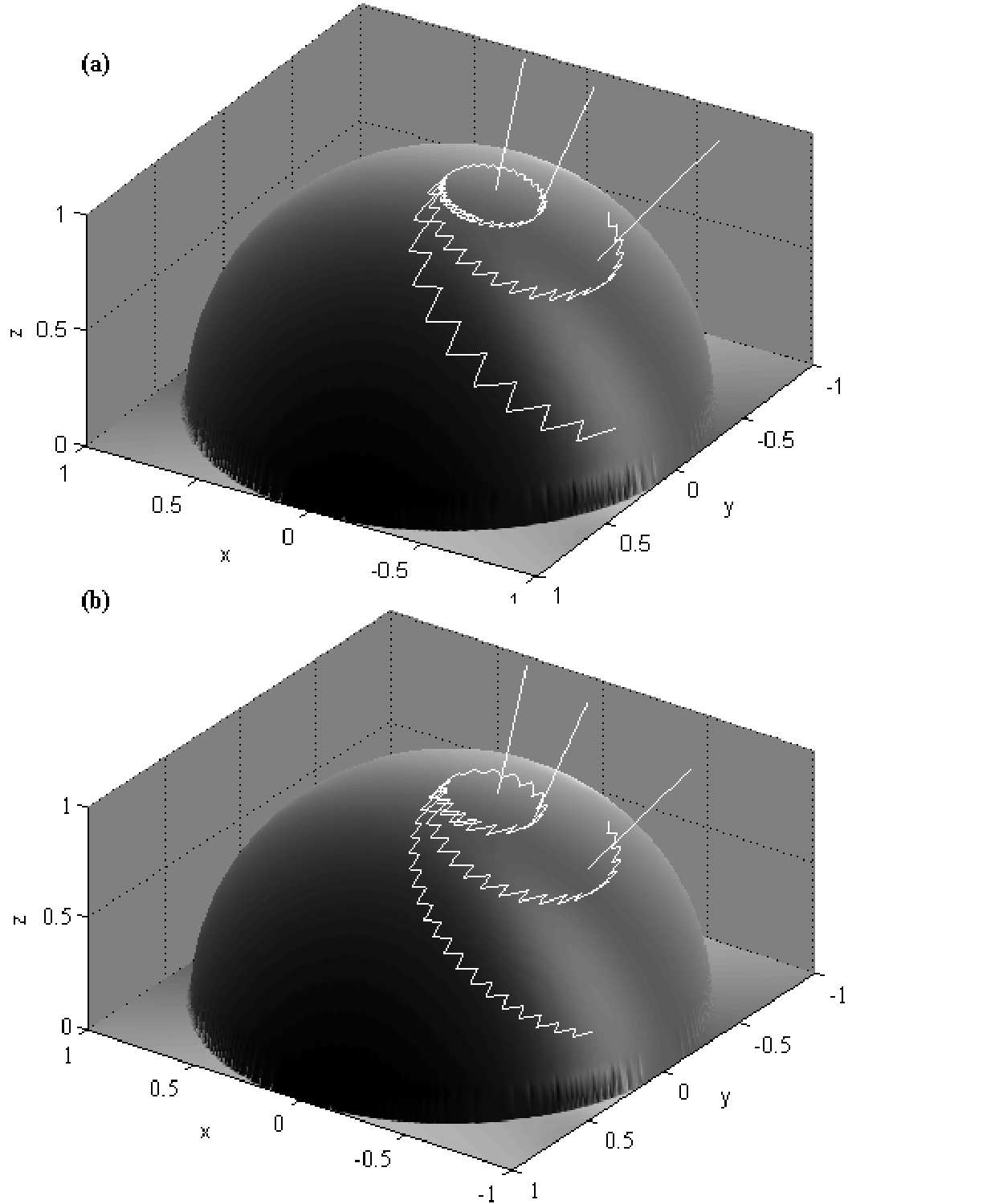}\\
  \caption{The curves at the unitary sphere give the evolution of the terminal
point of the vector $\r$ (represented with its initial point at the
origin) starting at $\r=\e_{3}$ and subjected to $N$ rotations
concerning $\R_{3}\left( \phi \right) \R_{1}\left( -\pi /2N\right)
=\R\left(\hat{n},\varphi\right)$. The line segments explicitate the
direction of each rotation axis $\hat{n}$. (a) $\phi =\pi /16$ and
three values for $N$: $N=10$, $N=20$ and $N=40$. Higher $N$
corresponds to $\hat{n}$ closer to $\e_{3}$ and to curves closer to
the initial point. (b) $N=20$ and three values for $\phi $: $\phi
=\pi /32$, $\phi =\pi /16$ and $\phi =\pi /8$. Higher $\phi$
corresponds to $\hat{n}$ closer to $\e_{3}$ and to curves closer to
the initial point.}
\end{figure}

For the initial vector  $\left|\psi(0)
  \right\rangle=|0_{a}\rangle|0_{b},1_{c}\rangle$, which corresponds to $(\r(0))^{T}=(0,0,1)$,
   the survival probability is

  \begin{equation}
P_{001}=|\r_{N}\cdot\r(0)|^2.
  \end{equation}

The structure of hamiltonians $H_{bc}$ and $H_{ab}$ conserves the
state vector in the subspace
$\{\left|1_{a},0_{b},0_{c}\right\rangle,
-i\left|0_{a},1_{b},0_{c}\right\rangle
,\left|0_{a},0_{b},1_{c}\right\rangle \}$, allowing us to assume the
presented geometrical interpretations. However, the control of
quantum state through interactions with a single auxiliary system is
not restricted to systems with such symmetry.

From a broader point of view, this effect occurs when the
accumulation of interactions with the same auxiliary subsystem
changes the signal and reduces the absolute value of the quantum
transition rate. In Fig.2 we show the function $\frac{dP_{001}}{dt}$
of the system $S_{a}-S_{b}-S_{c}$. The oscillations of
$\frac{dP_{001}}{dt}$ induce oscillations on the behavior of
$P_{001}(t)$ (increase-decrease). Therefore, for appropriate values
of the parameters ($N$ and interaction time with the auxiliary
subsystem) we may have the increase of the function $P_{001}(t)$, at
some time intervals, compensating the decrease in other time
intervals, inducing (in average) preservation of the initial state,
as it is shown in Fig.2.

\begin{figure}[h]
\centering
%\hspace{-3.5cm}
  \includegraphics[scale=0.5]{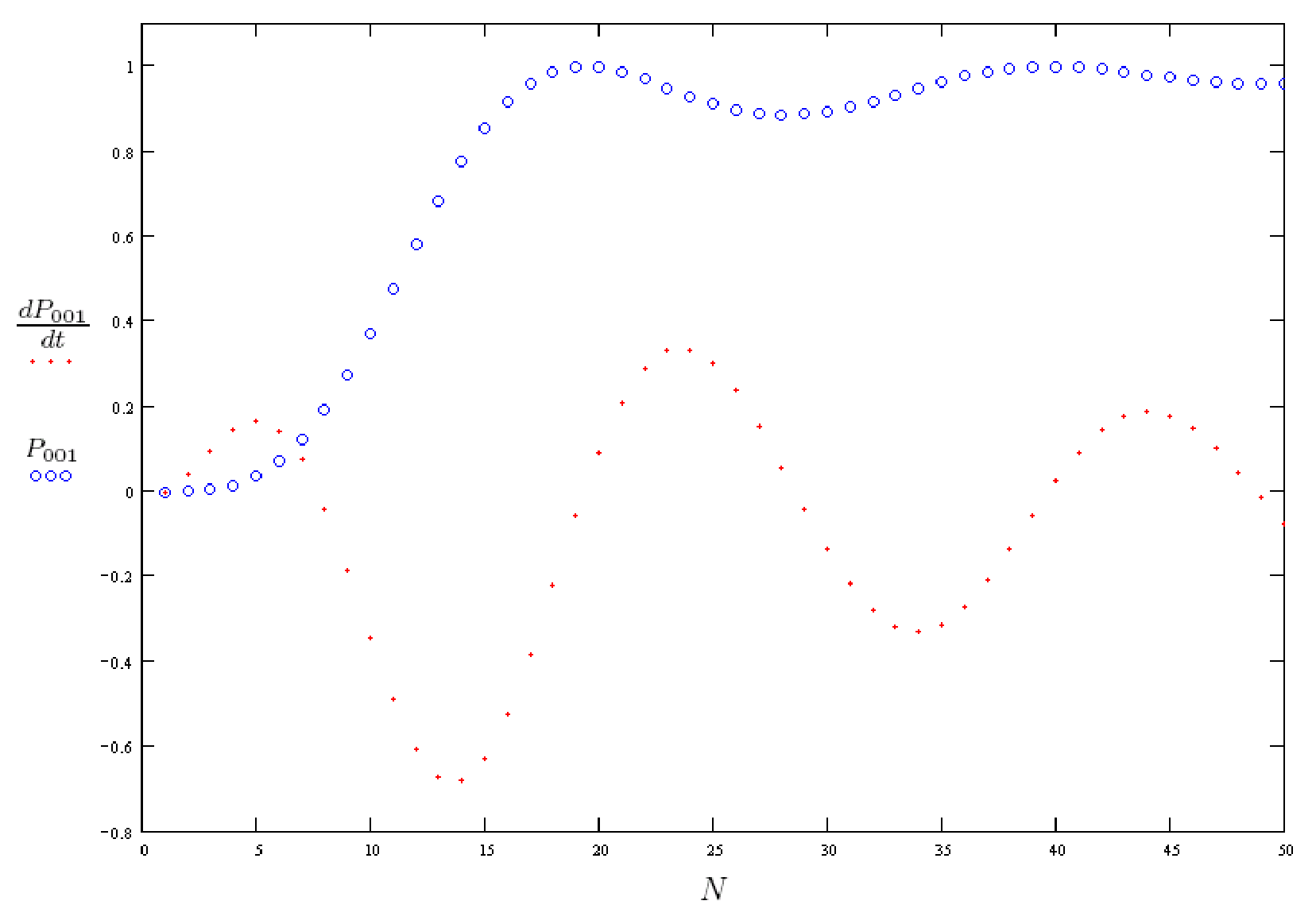}\\
  \caption{ Probability $P_{001}(t)$ and quantum transition rate $\frac{dP_{001}}{dt}(t)$ with $\phi=\frac{\pi}{10}$, $\theta=\frac{\pi}{2N}$}
\end{figure}

This inhibition of quantum transition induced by the increase on $N$
is similar to the discrete QZE, but is structured differently: the
QZE as presented in Ref. \cite{art12}, has a system of interest
interacting with $N$ auxiliary (probe) subsystems. After each
interaction the complete information about the occurrence of the
quantum transition is available on the probe. This fact implies the
cancelation of the transition rate after each interaction
\cite{art13}. Therefore, we may characterize the interactions
between system and probe as a measurement process (pre-measurement).
The net effect of $N$ interactions between \emph{a single auxiliary
system} and the system of interest can not be characterize as a
measurement process at all. Consequently for the later dynamics, the
quantum transition rate is not necessarily null after each
interaction, as it is shown in Fig.2.

Another difference between these two effects is that in the dynamics
presented here a transition like $|1_{a}\rangle|0_{b},0_{c}\rangle
\rightarrow|0_{a}\rangle|0_{b},1_{c}\rangle$ is intermediated by
$S_{b}$, \textit{i.e.}, if only one auxiliary subsystem interacts
$N$ times with the system of interest, the excitation present in
$S_{a}$ may return to the subsystem $S_{c}$ inducing on $P_{001}$
larger values than the survival probability observed in the QZE. A
comparison between these two probabilities is shown in Fig.3.

\begin{figure}[h]
\centering
%\hspace{-3.5cm}
  \includegraphics[scale=0.5]{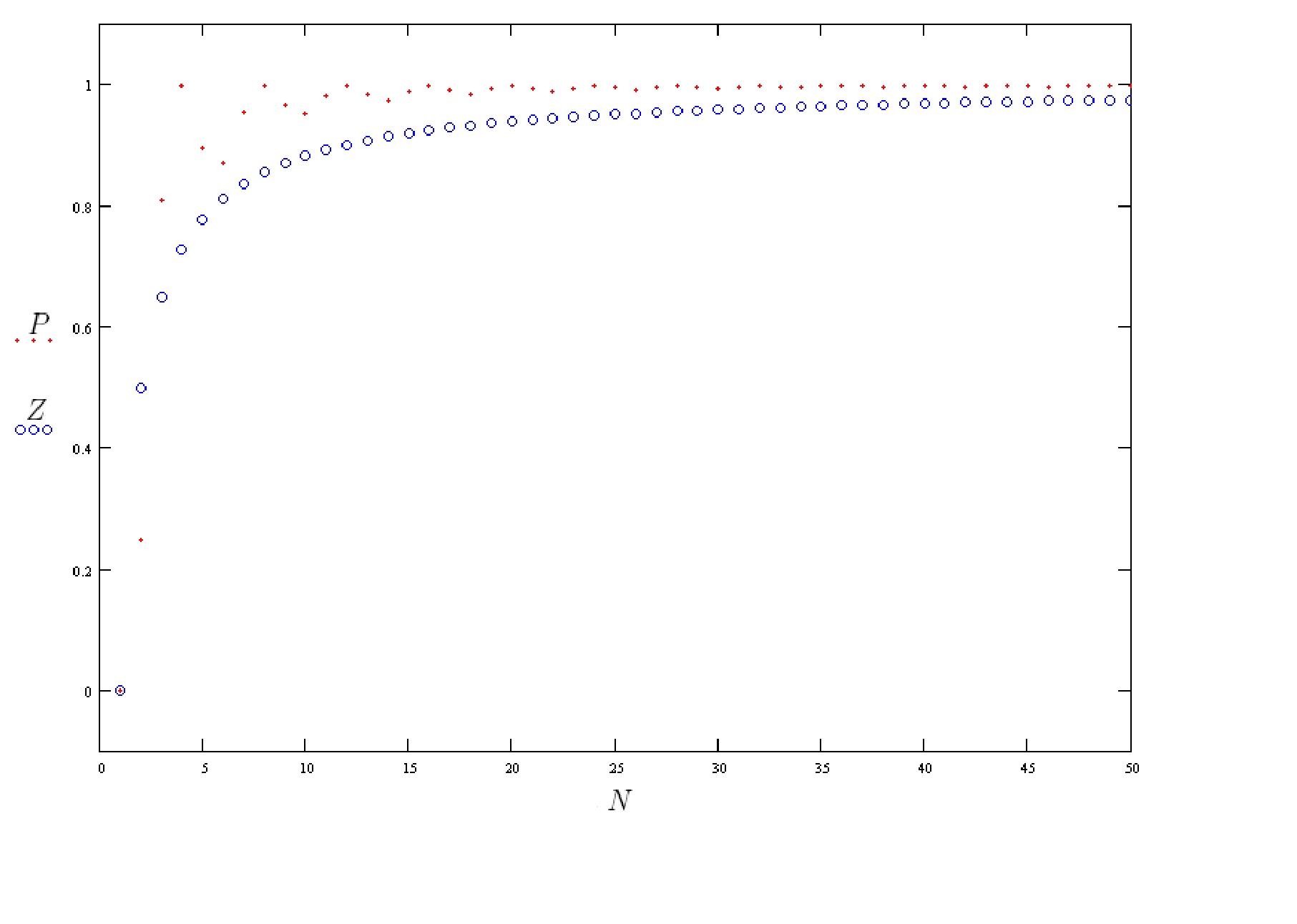}\\
  \caption{Comparison between the probability of non-transition related to QZE $Z(n)=\cos^{N}(\frac{\pi}{2N})$ and
  $P(n)=P_{001}$, with $\phi=\frac{\pi}{2}$ and $\theta=\frac{\pi}{2N}$}
\end{figure}

\section{Control of the entanglement dynamics}

Let us consider a system composed by two space-separated atoms, one
of them is isolated (atom $B$) and the other (atom $A$) is coupled
with an electromagnetic mode ($M_{1}$), as studied in Ref.
\cite{art10}. If there is an initial entanglement between atoms $B$
and $A$ (or between atom $B$ and mode $M_{1}$) it changes over time,
even if the initially entangled systems are not coupled. This
entanglement dynamics takes place because of the coupling between
atom $A$ and mode $M_{1}$. In this section we proposed a control of
this entanglement dynamics through successive interactions with a
\emph{single} auxiliary subsystem.

Suppose the initial state is

\begin{equation}
|\psi(0)\rangle=|g_{a}\rangle\left(\alpha|1_{1},g_{b}\rangle
+\beta|0_{1},e_{b}\rangle\right),\label{inicial}
\end{equation}
where the mode $M_{1}$ and the atom $B$ are entangled. The
coefficients $\alpha$ and $\beta$ give the intensity of the initial
entanglement.

The coupling between atom $A$ and $M_{1}$ can be described by the
Jaynes-Cummings model. After a time evolution the vector state of
the system has the form

\begin{equation}
|\psi(t)\rangle=\alpha\left(\cos(gt)|g_{a}\rangle|1_{1},g_{b}\rangle
-i\sin(gt)|e_{a}\rangle|0_{1},g_{b}\rangle\right)
+\beta|0_{1},e_{b}\rangle,\label{emt}
\end{equation}
where $g$ is the coupling coefficient between atom $A$ and $M_{1}$.
We consider the atomic transition frequency of atoms $A$ and $B$
resonant with the frequency of $M_{1}$. Notice that when
$t=\frac{\pi}{2}$ we have the entanglement swap, the entanglement
initially present in the subsystem $M_{1}$-atom $B$ is completely
transferred to atom $A$-atom $B$ subsystem.

To quantify and study the entanglement dynamics we write the
concurrence \cite{art14} between $M_{1}$ and atom $B$ (details of
this calculation are in Ref. \cite{art10}).

\begin{equation}
C_{M_{1},B}(t)=2|\alpha\beta\cos(gt)|.
\end{equation}
The concurrence oscillates assuming null values when
$gt=\frac{k\pi}{2}$, where $k$ is an odd number.

It is possible to control the entanglement dynamics through $N$
interactions with a single auxiliary subsystem. In the presentation
of such control we consider that $M_{1}$ is in a cavity that
supports two ortogonal modes ($M_{1}$ and $M_{2}$). An empirical
realization of this system (in a microwave cavity) is reported in
Ref. \cite{art11}.

The modes have different frequencies
($\omega_{1}-\omega_{2}=\delta$). The difference between atomic
energy levels may be controlled by Stark effect. The detuning
between the modes allows for the atom $A$, when coupled to $M_{1}$
($M_{2}$), not to interact with $M_{2}$ ($M_{1}$). Therefore it is
possible to control the coupling time between atom $A$ and the modes
$M_{1}$ and $M_{2}$.

A sequence of interactions between the atom $A$ and $M_{2}$,
inserted into the time evolution shown in (\ref{inicial}) and
(\ref{emt}), is responsible for the control on the dynamics of
entanglement between atom $B$ and $M_{1}$. The evolution of the
global system is composed by $N$ steps, each one in two stages. At
the first stage the atom $A$ interacts with $M_{1}$ and at the
second stage the atom $A$ interacts with $M_{2}$ (the second stage
is responsible for the control).

In the first stage the time evolution is governed by the hamiltonian

\begin{eqnarray}
H_{1}&=&\hbar\omega_{1}a_{1}^{\dagger}a_{1} +
\hbar\omega_{2}a_{2}^{\dagger}a_{2}
+\hbar\omega_{1}|e_{a}\rangle\langle
e_{a}|+\hbar\omega_{1}|e_{b}\rangle\langle e_{b}|+\hbar
g(\sigma_{a}^{+}a_{1} +
\sigma_{a}^{-}a_{1}^{\dagger}),\notag\\
&=&H^{'}_{1}+\hbar\omega_{1}|e_{b}\rangle\langle e_{b}|,\label{h1}
\end{eqnarray}
and in the second stage by the hamiltonian
\begin{eqnarray}
H_{2}&=&\hbar\omega_{1}a_{1}^{\dagger}a_{1} +
\hbar\omega_{2}a_{2}^{\dagger}a_{2}
+\hbar\omega_{2}|e_{a}\rangle\langle
e_{a}|+\hbar\omega_{1}|e_{b}\rangle\langle e_{b}|+\hbar
g(i\sigma_{a}^{+}a_{2} -i
\sigma_{a}^{-}a_{2}^{\dagger}),\notag\\
&=&H^{'}_{2}+\hbar\omega_{1}|e_{b}\rangle\langle e_{b}|,\label{h2}
\end{eqnarray}
where $\sigma_{k}^{+}=|e_{k}\rangle\langle g_{k}|$,
$\sigma_{k}^{-}=|g_{k}\rangle\langle e_{k}|$ ($k=a,b$). $H^{'}_{1}$
and $H^{'}_{2}$ act only on the subsystem composed by atom $A$,
$M_{1}$ and $M_{2}$. Notice that the coupling coefficient in the
second stage (in $H_{2}$) is imaginary, this is due to the ortogonal
mode's polarization.

The unitary time evolution operators for the first and second stages
are:

\begin{equation}
e^{-iH_{1(2)}t/\hbar}=e^{-iH^{'}_{1(2)}t/\hbar}e^{-i\omega_{1}|e_{b}\rangle\langle
e_{b}|t}.
\end{equation}

Written on the basis
$\{|0_{1},g_{a},1_{1}\rangle,|0_{1},e_{a},0_{1}\rangle,|1_{1},g_{a},0_{1}\rangle\}$
the operators $e^{-iH^{'}_{1}t/\hbar}$ and $e^{-iH^{'}_{2}t/\hbar}$
assume the form:

\begin{equation}
  e^{-iH^{'}_{1}t_{1}/\hbar} = e^{-i\omega_{1}t_{1}}\left[
    \begin{array}{ccc}
      e^{i\delta t_{1}} & 0 & 0 \\
      0 & \cos(gt_{1}) & -i\sin(gt_{1}) \\
      0 & -i\sin(gt_{1}) & \cos(gt_{1})
    \end{array} \right],\label{R1}
\end{equation}
\begin{equation}
  e^{-iH^{'}_{2}t_{2}/\hbar} = e^{-i\omega_{2}t_{2}}\left[
    \begin{array}{ccc}
      \cos(gt_{2}) & -\sin(gt_{2}) & 0 \\
      \sin(gt_{2}) & \cos(gt_{2})  & 0 \\
            0     &     0        & e^{-i\delta t_{2}}
    \end{array} \right],\label{R3}
\end{equation}

The time evolution is given by

\begin{equation}
|\psi_{N}\rangle=\left(e^{-iH^{'}_{2}t_{2}/\hbar}
e^{-iH^{'}_{1}t_{1}/\hbar}\right)^{N}|\psi(0)\rangle,
\end{equation}

and the initial vector state of the global system is

\begin{equation}
|\psi(0)\rangle=|g_{a}\rangle\left(\alpha|1_{1},g_{b}\rangle
+\beta|0_{1},e_{b}\rangle\right)|0_{2}\rangle,
\end{equation}
where $M_{2}$ (auxiliary subsystem) is prepared in the vacuum state.

After $N$ steps the interaction time between the atom $A$ and
$M_{1}$ is $T = Nt_{1}$. Let us consider $T = Nt_{1}=\frac{\pi}{2g}$
(as it is in Fig.4). When there is no participation of the auxiliary
subsystem in the dynamics the concurrence between the mode $M_{1}$
and the atom $B$ is null, because the entanglement is completely
transferred to the subsystem atom$A$-atom$B$. Therefore, with the
intervention of the auxiliary subsystem the entanglement dynamics is
inhibited, \textit{i.e.} the enhance on the number of interactions
with the auxiliary subsystem allows for the preservation of the
concurrence initial value, even when the total time for the
interaction of the system (atom $A$, atom $B$ and $M_{1}$) is $T =
Nt_{1}=\frac{\pi}{2g}$ (time of the entanglement swap), as it is
shown in Fig.4.

To summarize, we have presented a new strategy to control the
evolution of a quantum system. This strategy requires only unitary
interactions between the system of interest and a \emph{single
auxiliary subsystem}. We discuss two examples for the strategy
application. In the first one the auxiliary subsystem controls
excitation exchange between two qubits. In the second example, a
single auxiliary subsystem is used to control the entanglement of a
system composed by an isolated atom and a Jaynes-Cummings atom.

\begin{figure}[h]
\centering
%\hspace{-3.5cm}
  \includegraphics[scale=0.5]{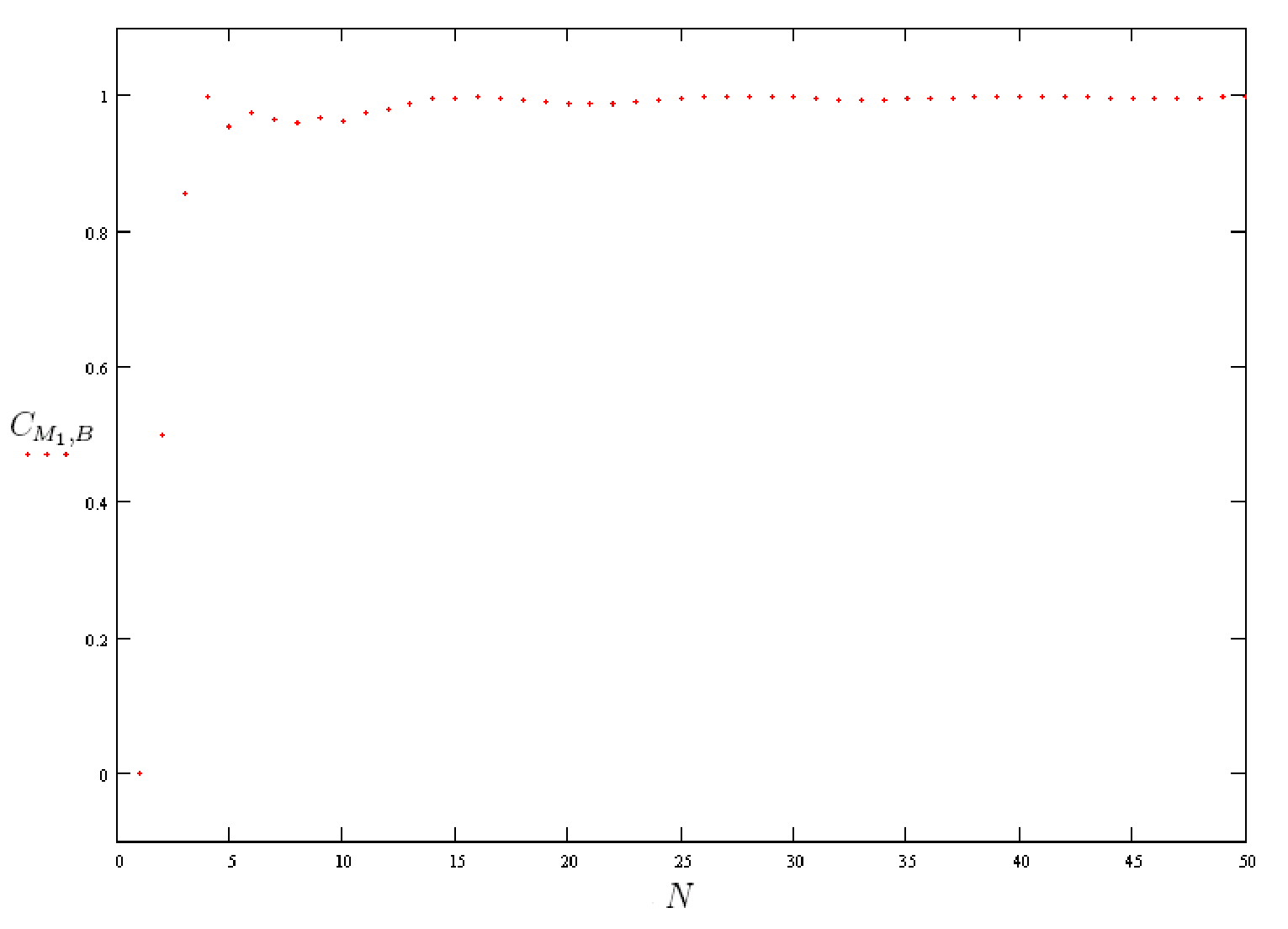}\\
  \caption{Concurrence between mode $M_{1}$ and atom $B$ with $\delta=8\times10^{5}s^{-1}$, $g=1.5\times10^{4}s^{-1}$ and $gt_{2}=\frac{\pi}{2}$.}
\end{figure}

\acknowledgments{M.C. Nemes and R. Rossi Jr. acknowledge financial
support by CNPq}

\appendix*
\section{Calculation of $\left| \psi_N\right\rangle$}

In this section we calculate explicitly the vector state $\left|
\psi_N\right\rangle$. Let us focus on the action of the unitary
matrix $\Uab(\phi)\Ubc(\theta)$ on the vector $\left|
\xi\right\rangle$ written in the basis
$\{\left|1_{a},0_{b},0_{c}\right\rangle,
-i\left|0_{a},1_{b},0_{c}\right\rangle
,\left|0_{a},0_{b},1_{c}\right\rangle \}$:

\begin{equation}
  \Uab(\phi)\Ubc(\theta)\left|
\xi\right\rangle =\Uab(\phi)\Ubc(\theta)\left[
    \begin{array}{c}
      \xi_1 \\
      \xi_2 \\
      \xi_3
    \end{array} \right] = \left[
    \begin{array}{c}
      \xi_1\cos\phi -\left(\xi_2\cos\theta+\xi_3\sin\theta\right)\cos\phi \\
      \xi_1\sin\phi+\left(\xi_2\cos\theta+\xi_3\sin\theta\right)\cos\phi \\
      \xi_3\cos\theta - \xi_2\sin\theta
    \end{array}\right]. \label{csifinal}
\end{equation}
This action may be mapped as a rotation on the real euclidian
subspace, by choosing conveniently the rotation matrix and assuming
that $\xi_1$, $\xi_2$ and $\xi_3$ (components of
$\left|\xi\right\rangle$) are real.

Defining

\begin{equation}
  \R_1\left(\varphi\right) \equiv \R\left(\e_1, \varphi\right) = \left[
    \begin{array}{ccc}
      1 & 0 & 0 \\
      0 & \cos\varphi & -\sin\varphi \\
      0 & \sin\varphi & \cos\varphi
    \end{array} \right],\label{R1}
\end{equation}
\begin{equation}
  \R_3\left(\varphi\right) \equiv \R\left(\e_3, \varphi\right) = \left[
    \begin{array}{ccc}
      \cos\varphi & -\sin\varphi & 0 \\
      \sin\varphi & \cos\varphi  & 0 \\
            0     &     0        & 1
    \end{array} \right],\label{R3}
\end{equation}
and the vector
\begin{equation}
  \r = \left[
    \begin{array}{c}
      \xi_1 \\
      \xi_2 \\
      \xi_3
    \end{array} \right],\label{r}
\end{equation}
we notice that the action of the matrix $\R_1\left(-\theta\right)$
and $\R_3\left(\phi\right)$, over $\r$ produces the same effect on
its components as the action of $\Ubc(\theta)$ and $\Uab(\phi)$ on
the components of $\left| \xi\right\rangle$, \textit{i.e.},
\begin{equation}
  \R_3\left(\phi\right)\R_1\left(-\theta\right)\r= \left[
      \begin{array}{c}
        \xi_1\cos\phi - \left(\xi_2\cos\theta + \xi_3\sin\theta\right)\sin\phi \\
        \xi_1\sin\phi + \left(\xi_2\cos\theta + \xi_3\sin\theta\right)\cos\phi \\
        \xi_3\cos\theta - \xi_2\sin\theta
      \end{array}\right]. \label{rfinal}
\end{equation}

The ortogonal matrix $\R_3\left(\phi\right)\R_1\left(-\theta\right)$
may be written as
\begin{equation}
  \R_3\left(\phi\right)\R_1\left(-\theta\right) = \R\left(\n,\varphi\right)=\exp\left(\varphi \n\cdot \J\right),
  \label{exp}
\end{equation}
where $\n\cdot \J$ is the generator of rotations around the axis
defined by the unitary vector $\n = a\e_1 + b\e_2 +c\e_3$, $\varphi$
is the angle of rotation around $\n$, $\J = \e_1 \mathbf{J}_1 + \e_2
\mathbf{J}_2 + \e_3 \mathbf{J}_3$ where $\mathbf{J}_1$,
$\mathbf{J}_2 $ and $\mathbf{J}_3$ are the generators of rotation
around the axis $Ox$, $Oy$ and $Oz$, respectively.
\begin{equation}
  \mathbf{J}_1 = \left[
    \begin{array}{ccc}
      0 & 0 & 0 \\
      0 & 0 & -1 \\
      0 & 1 & 0
    \end{array} \right],\:
  \mathbf{J}_2 = \left[
    \begin{array}{ccc}
      0 & 0 & 1 \\
      0 & 0 & 0 \\
      -1 & 0 & 0
    \end{array} \right],\:
  \mathbf{J}_3 = \left[
    \begin{array}{ccc}
      0 & -1 & 0 \\
      1 & 0 & 0 \\
      0 & 0 & 0
    \end{array} \right].\label{jotas}
\end{equation}

In order to calculate the action of
$\left(\Uab(\phi)\Ubc(\theta)\right)^N$ over the state vector
$\left|\psi_{0}\right\rangle$, we use the mapping of the operator
$\Uab(\phi)\Ubc(\theta)$ on the ortogonal matrix
$\R_3\left(\phi\right)\R_1\left(-\theta\right)$ and the identity:

\begin{equation}
    \left[\R_3\left(\phi\right)\R_1\left(-\theta\right)\right]^N =
    \left[\exp\left(\varphi \n\cdot \J\right)\right]^N =\exp\left(N\varphi \n \cdot\J\right).
    \label{final}
\end{equation}

The calculation of
$\left[\R_3\left(\phi\right)\R_1\left(-\theta\right)\right]^N$ is
reduced now to finding the axis $\n$ and the angle $\varphi$.

\subsubsection{Finding $\n$ and $\varphi$}

Explicitly,
\begin{equation}
  \varphi\n\cdot\J = \varphi\left[
    \begin{array}{ccc}
      0  &    -c   & b  \\
      c  &     0   & -a \\
      -b &     a   &  0
    \end{array} \right].\label{ma1}
\end{equation}
After some algebra we have
\begin{equation}
  \left(\varphi\n\cdot\J\right)^{2n} = \varphi^{2n}\left(-1\right)^{n+1}
  \left(\n\cdot\J\right)^{2}, \:
  \left(\varphi\n\cdot\J\right)^{2n+1} = \varphi^{2n+1}\left(-1\right)^{n}
  \left(\varphi\n\cdot\J\right),
  \label{potenciasJ}
\end{equation}
and we may write
\begin{equation}
  \exp\left(\varphi \n\cdot \J\right) = \mathbf{1} + \left(1-\cos\varphi \right)
  \left(\n\cdot\J\right)^2 + \sin\varphi\left(\n\cdot\J\right).
  \label{com_modulo}
\end{equation}

The calculation of $\left(\n\cdot\J\right)^2$ is given by:
\begin{equation}
  \left(\n\cdot\J\right)^2 =  \left[
    \begin{array}{ccc}
        a^2 - 1  &   ab     &   ac \\
          ab     &  b^2 - 1 &   bc \\
          ac     &   bc     & c^2 - 1
    \end{array} \right]. \label{ma2}
\end{equation}

Substituting equations (\ref{ma1}) and (\ref{ma2}) in
(\ref{com_modulo}) we get

\begin{equation}
  \exp\left(\varphi \n\cdot \J\right)  =  \left[
    \begin{array}{ccc}
      \left(1-a^2\right)\cos\varphi + a^2 & ab\left(1-\cos\varphi\right)-c\sin\varphi & ac\left(1-\cos\varphi\right)+b\sin\varphi \\
      ab\left(1-\cos\varphi\right)+c\sin\varphi & \left(1-b^2\right)\cos\varphi + b^2 & bc\left(1-\cos\varphi\right)-a\sin\varphi \\
      ac\left(1-\cos\varphi\right)-b\sin\varphi & bc\left(1-\cos\varphi\right)+a\sin\varphi & \left(1-c^2\right)\cos\varphi + c^2
    \end{array} \right].
  \label{expJ}
\end{equation}

Comparing (\ref{expJ}) with the product
\begin{equation}
  \R_3\left(\phi\right)\R_1\left(-\theta\right)  =  \left[
    \begin{array}{ccc}
      \cos\phi    &   -\sin\phi \cos\theta   &  -\sin\phi \sin\theta   \\
      \sin\phi    &    \cos\phi \cos\theta   &   \cos\phi \sin\theta   \\
         0        &      -\sin\theta         &       \cos\theta
    \end{array} \right],
  \label{R3R1}
\end{equation}
we get the following expressions for $\sin\varphi$, $\cos\varphi$
and components of $\n$:
\begin{subequations}
    \label{seno-cosseno}
    \begin{align}
        \sin\varphi &= 2\cos\frac{\phi}{2}\cos\frac{\theta}{2}\sqrt{\sin^2\frac{\phi}{2}
        +\sin^2\frac{\theta}{2} - \sin^2\frac{\theta}{2}\sin^2\frac{\phi}{2}},
        \label{seno}\\
        \cos\varphi &=\frac{\cos\phi +\cos\theta+\cos\phi\cos\theta
        -1}{2},
        \label{cosseno}
    \end{align}
\end{subequations}
\begin{subequations}
    \label{eixo}
    \begin{align}
        a &= -\frac{\sin\theta\left(\cos\phi + 1\right)}
        {4\cos\frac{\phi}{2}\cos\frac{\theta}{2}\sqrt{\sin^2\frac{\phi}{2}
    +\sin^2\frac{\theta}{2} - \sin^2\frac{\theta}{2}\sin^2\frac{\phi}{2}}}, \label{a} \\
        b &= -\frac{\sin\phi\sin\theta}
        {4\cos\frac{\phi}{2}\cos\frac{\theta}{2}\sqrt{\sin^2\frac{\phi}{2}
    +\sin^2\frac{\theta}{2} - \sin^2\frac{\theta}{2}\sin^2\frac{\phi}{2}}}, \label{b} \\
        c &= \frac{\sin\phi\left(\cos\theta + 1\right)}
        {4\cos\frac{\phi}{2}\cos\frac{\theta}{2}\sqrt{\sin^2\frac{\phi}{2}
    +\sin^2\frac{\theta}{2} - \sin^2\frac{\theta}{2}\sin^2\frac{\phi}{2}}}. \label{c}
    \end{align}
\end{subequations}

This result allows us to calculate
$\left[\R_3\left(\phi\right)\R_1\left(-\theta\right) \right]^N$.
\begin{equation}
    \left[\R_3\left(\phi\right)\R_1\left(-\theta\right)\right]^N =
    \left[\exp\left(\varphi \n \cdot\J\right)\right]^N =\exp\left(N\varphi \n \cdot\J\right).
    \label{final}
\end{equation}

The exponential $\exp\left(N\varphi \n \cdot\J\right)$ will have the
form identical to the matrix \eqref{expJ}, but with the substitution
of $\varphi$ by $N\varphi$. The components $a$, $b$ and $c$, as well
as $\sin\varphi$ and $\cos\varphi$ are shown in \eqref{seno-cosseno}
and \eqref{eixo}.


\begin{thebibliography}{14}



\bibitem{art1} M. A. Nielsen and  I. L. Chuang, {\it Quantum Computation
and Quantum Information} (Cambridge University Press, Cambridge,
England, 2000).



\bibitem{art2} Y. P. Huang and M. G. Moore, \pra \textbf{77}, 062332 (2008).



\bibitem{art3} J. D. Franson, B. C. Jacobs, and T. B. Pittman, \pra \textbf{70}, 062302 (2004).


\bibitem{art4} S. Maniscalco, F. Francica, R. L. Zaffino, N. Lo Gullo and F. Plastina,
\prl \textbf{100}, 090503 (2008).


\bibitem{art5} D. Dhar, L. K. Grover, and S. M. Roy, \prl \textbf{96}, 100405 (2006).


\bibitem{art6} P. Facchi and S. Pascazio, \prl \textbf{89}, 080401
(2002).


\bibitem{art7} A. Beige, D. Braun, B. Tregenna, and P. L. Knight,
\prl \textbf{85}, 1762 (2000).



\bibitem{art8} L. Viola and S. Lloyd, \pra \textbf{58}, 2733 (1998).



\bibitem{art9} D. Rossini, P. Facchi, R. Fazio, G. Florio, D. A. Lidar, S. Pascazio, F. Plastina, and P.
Zanardi, \pra \textbf{77}, 052112 (2008).



\bibitem{art10} Zhi-Jian Li, Jun-Qi Li, Yan-Hong Jin and Yi-Hang Nie, J. Phys. B \textbf{40}, 3401 (2007).


\bibitem{art11} A. Rauschenbeutel, P. Bertet, S. Osnaghi, G. Nogues, M. Brune, J. M. Raimond, and S. Haroche,
 \pra \textbf{64}, 050301  (2001).


\bibitem{art12} S. Pascazio, M. Namiki, \pra \textbf{50}, 4582 (1994).



\bibitem{art13} A. F. R. de Toledo Piza and M. C. Nemes, Physics Letters A \textbf{290}, 6 (2001).



\bibitem{art14} S. Hill and W. K. Wootters, \prl \textbf{78}, 5022 (1997).



\end{thebibliography}
\end{document}